\documentclass[fleqn,usenatbib,useAMS]{mnras}
\usepackage[table,xcdraw]{xcolor}
\usepackage{newtxtext}
\usepackage[utf8]{inputenc}
\usepackage[english]{babel}
\usepackage[T1]{fontenc}
\usepackage{color}
\usepackage{deluxetable}
\usepackage{amssymb,amsmath}
\usepackage{rotating}
\usepackage{graphicx}
\usepackage{hyperref}
\hypersetup{colorlinks,citecolor=blue,filecolor=blue,linkcolor=blue,urlcolor=blue}
\usepackage{mathtools}
\usepackage{mathpazo}
\usepackage{graphicx}
\usepackage{footnote}
\usepackage{longtable}
\usepackage{soul}
\usepackage{caption}
\usepackage{natbib}
\title[INOV of RQNLSy1s] {Intra-night optical variability of peculiar narrow-line Seyfert 1 galaxies with enigmatic jet behavior}

\author[Ojha Vineet, VS, MB, EJ ]{Vineet Ojha\thanks{E-mail:vineetojhabhu@gmail.com, Current affiliation: Kavli Institute for Astronomy and Astrophysics, Peking University, Beijing \it{100871}, China}$^{1}$,  Veeresh Singh$^{1}$, M. Berton$^{2}$, E. J{\"a}rvel{\"a}$^{3}$ \\
  $^{1}$Physical Research Laboratory (PRL), Astronomy and Astrophysics Division, Ahmedabad, \it{380 009}; India \\
  $^{2}$European Southern Observatory (ESO), Alonso de Córdova 3107, Casilla 19, Santiago \it{19001}, Chile\\
 $^{3}$Homer L. Dodge Department of Physics and Astronomy, The University of Oklahoma, 440 W. Brooks St., Norman, OK \it{73019}, USA}
 
\date{Accepted XXX. Received YYY; in original form ZZZ}
\pubyear{2022}

\begin{document}
\label{firstpage}
\pagerange{\pageref{firstpage}-- \pageref{lastpage}}
\maketitle
\begin{abstract}
Variability studies of active galactic nuclei are a powerful diagnostic tool in understanding the physical processes occurring in disk-jet regions, unresolved by direct imaging with currently available techniques. Here, we report the first attempt to systematically characterize intra-night optical variability (INOV) for a sample of seven apparently radio-quiet narrow-line Seyfert 1 galaxies (RQNLSy1s) that had shown recurring flaring at 37 GHz in the radio observations at Mets{\"a}hovi Radio Observatory (MRO), indicating the presence of relativistic jets in them, but no evidence for relativistic jets in the recent radio observations of Karl G. Jansky Very Large Array (JVLA) at 1.6, 5.2, and 9.0 GHz. We have conducted a total of 28 intra-night sessions, each lasting $\geq$ 3 hrs for this sample, resulting in an INOV duty cycle ($\overline{DC} ~\sim$20\%) similar to that reported for $\gamma$-ray-NLSy1s (DC $\sim$25\% - 30\%), that display blazar-like INOV. This in turn infers the presence of relativistic jet in our sample sources. Thus, it appears that even lower-mass (M$_{BH} \sim$10$^{6}$ M$_{\sun}$) RQNLSy1 galaxies can maintain blazar-like activities. However, we note that the magnetic reconnection in the magnetosphere of the black hole can also be a viable mechanism to give rise to the INOV from these sources. 


\end{abstract}



\section{Introduction}
 \label{sec1.0}
Accretion of gas around the central supermassive black holes (SMBH) of masses $M_{SMBH}\sim10^{6} - 10^{10} M_{\sun}$ is the main powering mechanism of active galactic nuclei (AGNs) that makes them the most energetic objects in the universe with integrated luminosities reaching up to 10$^{48}$ erg s$^{-1}$~\citep{Koratkar1999PASP..111....1K, Bischetti2017A&A...598A.122B}. Among the different observational characteristics, variability on different timescales ranging from minutes to decades across the electromagnetic spectrum is being used as one of the defining characteristics of AGNs~\citep{Gaskell2003A&AT...22..661G, Padovani2017A&ARv..25....2P}. Variability studies in AGNs play an important role in understanding the physical processes occurring in these objects. For instance,  AGN variability has been used to probe emission mechanisms occurring on physical scales that are unresolved by currently available telescopes/facilities, and also used to investigate the spin and mass of the central SMBH~\citep{Urry1995PASP..107..803U, Wagner1995ARA&A..33..163W, Ulrich1997ARA&A..35..445U, Zensus1997ARA&A..35..607Z, Cackett2013ApJ...764L...9C, McHardy2014MNRAS.444.1469M, Emmanoulopoulos2014MNRAS.439.3931E}. It is widely believed that the optically thick accretion disk surrounding the SMBH is primarily responsible for the optical emission from AGNs~\citep{Shakura1973A&A....24..337S}, but the physical processes producing optical variability are not clearly understood.  In a tiny subset of AGNs called blazars, the optical variability is thought to originate due to relativistic boosting of small fluctuations arising through the turbulence of plasma in the jet~\citep[e.g., see][]{Marscher1991vagn.conf..153M, Goyal2012A&A...544A..37G, Calafut2015JApA...36..255C}. Short-term optical flux variability of AGNs from minutes to hours is known as `Intra-Night Optical Variability'~\citep[INOV, ][]{Gopal-Krishna2003ApJ...586L..25G}.\par
Among the INOV study of different luminous classes covered by~\citet{Goyal2013MNRAS.435.1300G}, strong INOV with duty cycle (DC) above 30 percent is exhibited by high optical polarization core-dominated quasars, and TeV blazars (both are radio-loud\footnote{\small Radio-loudness is parameterized by the ratio of radio
to optical flux densities at  5 GHz and at 4400\AA, respectively, with RL$\leq$ 10 and $>$ 10 for radio-quiet and radio-loud AGNs, respectively~\citep[e.g. see,][]{Kellermann1989AJ.....98.1195K}.}).
This suggests that strong INOV can be an effective tracer of jet activity in AGNs. On the other hand, a low level of INOV DC $<$ 10 percent observed in radio-quiet quasars~\citep[e.g., see][]{Goyal2013MNRAS.435.1300G} is thought to originate either from its weak jet~\citep[e.g.,][]{Kellermann2016ApJ...831..168K} and/or transient shocks or `hot spots' in the accretion disk around the SMBH~\citep{Chakrabarti1993ApJ...411..602C, Mangalam1993ApJ...406..420M}.\par
Among the different sub-classes of lower-luminosity AGNs, a handful of the radio-loud minority of Narrow-line Seyfert 1 (NLSy1) galaxies, marked by detection in $\gamma$-ray band, exhibit comparable DC ($\sim$ 30 percent) to the jetted class of AGN~\citep[e.g., see][]{Paliya2013MNRAS.428.2450P, Ojha2021MNRAS.501.4110O}. NLSy1s are characterized by smaller width of Balmer emission lines with FWHM(H${\beta}) < $ 2000 km s$^{-1}$ and flux ratio of [O$_{III}]_{\lambda5007}/H\beta$ $<$ 3~\citep{Osterbrock1985ApJ...297..166O, Shuder-Osterbrock1981ApJ...250...55S, Goodrich1989ApJ...342..908G}. The ample majority of NLSy1s are radio-quiet, only a tiny fraction $\sim$ 7\% is radio-loud and likely to harbor relativistic jets~\citep{Komossa2006AJ....132..531K, Singh2018MNRAS.480.1796S}. The presence of relativistic jets has been proven in gamma-ray-detected RLNLSy1s~\citep[$\gamma$-RLNLSy1s, e.g.,][]{Giroletti2011A&A...528L..11G, Caccianiga2015MNRAS.451.1795C, Jarvela2022A&A...658A..12J}. Unlike the general population of RLNLSy1s, $\gamma$-RLNLSy1s show much stronger INOV similar to blazars, suggesting a connection of relativistic jets with their strong INOV. Despite many similar characteristics such as flat radio spectrum, high brightness temperature, superluminal motion, and $\gamma$-ray detection~\citep{Yuan2008ApJ...685..801Y, Abdo2009ApJ...699..976A, Abdo2009ApJ...707..727A, Abdo2009ApJ...707L.142A, Foschini2010ASPC..427..243F, Foschini2011nlsg.confE..24F, D'Ammando2012MNRAS.426..317D, Berton2017FrASS...4....8B, D'Ammando2015MNRAS.452..520D, Yao2015MNRAS.454L..16Y,  Lister2018rnls.confE..22L, Paliya2018ApJ...853L...2P, Yang2018MNRAS.477.5127Y, Yao2019MNRAS.487L..40Y, Foschini2021Univ....7..372F, Foschini2022Univ....8..587F, Li2023A&A...676A...9L} to blazar class of AGN, an important difference between NLSy1s and blazars is an order of less massive black hole for NLSy1s~\citep[$10^{7} M_{\sun}$, ][]{Grupe2004ApJ...606L..41G, Deo2006AJ....132..321D, Peterson2011nlsg.confE..32P}. This makes them to harbor less powerful jets~\citep{Angelakis2015A&A...575A..55A, Gu2015ApJS..221....3G, Fuhrmann2016RAA....16..176F, Paliya2019JApA...40...39P}.\par
Recently,  a sample of seven radio-quiet and/or radio-silent (never detected in radio) NLSy1s exhibited recurring flaring at 37 GHz in the radio observations at MRO~\citep[see][]{2018A&A...614L...1L}.  However, when these RQNLSy1s were observed with JVLA in A configuration at three different frequencies, 1.6 GHz, 5.2 GHz, and 9.0 GHz, no hints of relativistic jet was detected from them~\citep[see ][]{Berton2020A&A...636A..64B}. Taking advantage of INOV, here, we investigate the presence of relativistic jets in these NLSy1s by performing intranight optical monitoring of these seven RQNLSy1s with 1$-$2.5m ground-based optical telescopes.\par
The format of this paper is as follows. In Sect.~\ref{section_2.0}, we describe our optical monitoring and data reduction procedure. Sect.~\ref{sec_3.0} provides details of statistical analysis methods. The main results of this work followed by discussion are given in Sect.~\ref{sec_4.0}.

\begin{table}
 \caption{A current sample of 7 RQNLSy1s.}
\label{tab:source_info}
\begin{tabular}{lllll}
 \hline
\multicolumn{1}{l}
  SDSS name$^{a}$ &   R-mag$^{b}$  &  $z^{c}$  & $RL^{d}$ & $log (M_{BH})^{e}$   \\     
                    &               &          &                    & \multicolumn{1}{r} {M$_{\sun}$} \\
\hline
J102906.69$+$555625.2    & 19.10 & 0.45 & --   & 7.33  \\
J122844.81$+$501751.2    & 17.80 & 0.26 & --   & 6.84\\
J123220.11$+$495721.8    & 16.90 & 0.26 & --   & 7.30 \\
J150916.18$+$613716.7    & 18.60 & 0.20 & --   & 6.66\\
J151020.06$+$554722.0    & 17.80 & 0.15 & --   & 6.67\\
J152205.41$+$393441.3    & 13.10 & 0.08 & 02   & 5.97\\
J164100.10$+$345452.7    & 16.00 & 0.16 & 13   & 7.15\\

\hline
\multicolumn{5}{l}{$^{a}$SDSS name of RQNLSy1s.}\\
\multicolumn{5}{l}{$^{b}$R-band magnitude of NLSy1s taken from~\citet{Monet1998AAS...19312003M}.}\\
\multicolumn{5}{l}{$^{c}$Redshift of the RQNLSy1s taken from {~\citet{2018A&A...614L...1L}}}.\\
\multicolumn{5}{l}{$^{d,~e}$Both radio-loudness or radio-silent $RL\equiv S_{1.4~GHz}/S_{440~nm}$} \\
\multicolumn{5}{l}{and black hole masses of current sources are taken from} \\ 
\multicolumn{5}{l}{{\citet{2015A&A...573A..76J}} and{~\citet{2018A&A...614L...1L}}. In both}\\ \multicolumn{5}{l}{articles, black hole masses were estimated following the }\\
\multicolumn{5}{l}{FWHM(H$\beta$) - luminosity mass scaling relation, given}\\
 \multicolumn{5}{l}{by~\citet{Greene2005ApJ...630..122G}.}\\

\end{tabular}
\end{table}

\begin{table*}
 \begin{center}
  \caption{Details of system parameters of the telescopes and detectors used in the observations of 7 RQNLSy1s. }
  \label{tab:telescope_info}
  \begin{tabular}{c c c c c c c c c c}
		\hline 
	Telescope (s) 	& No. of sessions	   & Detector (s)     & Readout speed&  \multicolumn{1}{l}{Field of view}    & Readout  & Gain        & Focal ratio & Pixel size & Plate scale\\
		                   &        &          & &(arcmin$^{2}$)     & noise     & (e$^-$ & of & of CCD     & of CCD\\
                              &     &                  &                  &  &(e$^-$)              &    /ADU)         &     telescope   & ($\mu$m) & ( $^{\prime\prime}$/pixel )   \\       
\hline
 1.04m ST$^{a}$      & 05   & 4k$\times$4k    & 100 kHz & 15.70$\times$15.70 & 3.0     &10.0 & f/13  & 15.0 & 0.23  \\
 1.30m DFOT$^{b}$    & 10   & 2k$\times$2k    & 100 kHz & 18.27$\times$18.27& 7.5     & 2.0 & f/4   & 13.5 & 0.53 \\
 1.20m Mt Abu$^{c}$  & 12   & 1k$\times$1k    &  50 kHz & 5.21$\times$5.21& 5.0     & 5.0 & f/13   & 13.0 & 0.30 \\ 
 2.50m Mt Abu$^{c}$  & 01   & 4k$\times$4k    & 100 kHz & 10.00$\times$10.00& 2.1     & 3.0 & f/8   & 15.0 & 0.15 \\ 
 	
		\hline

 \multicolumn{8}{l}{$^{a}$Sampurnand Telescope (ST), $^{b}$Devasthal Fast Optical Telescope (DFOT), $^{c}$Mount Abu telescope.}\\
                 
  \end{tabular}
\end{center}

\end{table*}

\section{Optical intra-night monitoring and data reduction}
\label{section_2.0}
Optical telescopes from two Indian institutes namely Aryabhatta Research Institute of Observational Sciences (ARIES) and Physical Research Laboratory (PRL) were used for the Intra-night monitoring of the seven RQNLSy1s in the broad-band Johnson-Cousin filter R except for a session with 2.5m PRL telescope when it was taken in the SDSS filter r due to non-availability of broad-band Johnson-Cousin filter R. Broad-band Johnson-Cousin filter R and SDSS filter r were chosen for observations because CCD detector used has maximum sensitivity in these bands. A total of four telescopes two from ARIES namely 1.04 meter (m) Sampurnanand telescope~\citep[ST,][]{Sagar1999Csi...77...77.643}, 1.30m Devasthal Fast Optical Telescope~\citep[DFOT,][]{Sagar2010ASInC...1..203S} located at Nainital, Uttarakhand, and two from PRL namely, 1.2m telescope~\citep{Srivastava2021ExA....51..345S} and 2.5m telescope located at Mount Abu Rajasthan, were used in this work. The details of the observational set-ups used for each telescope in observing the sample of 7 RQNLSy1s are listed in Table~\ref{tab:telescope_info}. 
At least three epochs of observation each $\geq$ 3 hours were devoted for each RQNLSy1s. Since we have used in the present work, the optical telescopes range between 1.04m and 2.5m, therefore, the typical exposure time was set between 300 sec to 1200 sec to reach a suitable signal-to-noise ratio (SNR), depending on the sky condition, moon phase, telescope efficiency, and magnitude (brightness) of the RQNLSy1s.\par
For preliminary processing of the raw images, at least three bias frames, and also three flat frames were taken during each observing session. Furthermore, the standard tasks available in the {\sc IRAF}\footnote{Image Reduction and Analysis Facility (http://iraf.noao.edu/)} software package were followed for making final science images from the raw images. Since the field of each target RQNLSy1s was not clustered, therefore, aperture photometry~\citep{1987PASP...99..191S, 1992ASPC...25..297S} was used in the current work for extracting the instrumental magnitudes of RQNLSy1s and the comparison stars registered in the CCD frames, using DAOPHOT II algorithm\footnote{Dominion Astrophysical Observatory Photometry (http://www.astro.wisc.edu/sirtf/daophot2.pdf)}. As emphasized in~\citet{Ojha2021MNRAS.501.4110O} size of the chosen aperture is an important parameter while estimating the instrumental magnitude and the corresponding SNR of the individual photometric data points registered on the CCD frames. Additionally, caution about the point spread function (PSF) variation becomes very important when dealing with intra-night variability of nearby ($\leq$ 0.4) AGNs. Because in such a situation a significant contribution to the total flux can come from the underlying host galaxy that can mimic the INOV in the standard analysis of the differential light curves (DLCs) due to the significant relative contributions of the (point-like) AGN and the host galaxy to the aperture photometry with the variation of PSF during the session~\citep{Cellone2000AJ....119.1534C}. Therefore, the procedure of data reduction, PSF estimation for aperture photometry, selection of aperture, and caution for PSF variations (see Sect.~\ref{sec_4.0}) were followed from~\citet{Ojha2021MNRAS.501.4110O}. Since except for one RQNLSy1 J102906.69$+$555625.2~($z = 0.45$), all the RQNLSy1s in the present sample are at lower redshift ($\leq$ 0.4), therefore proper caution has been taken about its PSF variation during the night before commenting about its variability (see Sect.~\ref{sec_3.0}).\par
Furthermore, DLCs of target RQNLSy1 for each intra-night session were derived relative to a pair of non-varying (steady) comparison stars (see online Figs. A1), additionally, the PSF variation during each intra-night session is plotted in the bottom panel of DLCs.

\section{Methodology for analysis}
\label{sec_3.0}

\subsection{Statistical tests}
To examine the presence of intra-night variability in the present sample of 7 RQNLSy1s, we have applied two different flavors of F-test,  (i) standard \emph{$F$-test} ($F^{\eta}$-test) and (ii) the power-enhanced \emph{$F$-test} ($F_{p-enh}$-test), by following the basic requirement and procedure of these two tests as described in~\citep{Goyal2012A&A...544A..37G, Diego2014AJ....148...93D}. Several star$-$star DLCs were generated for each session with the instrumental magnitudes extracted from the aperture photometry, and out of several pairs of star$-$star DLCs those two stars were chosen as comparison stars for which no-variability resulted based upon $F^{\eta}$-test. Out of the chosen two comparison stars, the one with the closest match ($\Delta{m} \sim $1, a requirement of $F^{\eta}$-test) in magnitude to the target RQNLSy1 is chosen as a reference star, and other as comparison star. Furthermore, two versions of \emph{$F$-test} are applied to the DLCs of target RQNLSy1 relative to the reference star and comparison star (basic parameters of these two stars are tabulated in the online Table A1). The $F^{\eta}$-values for the two RQNLSy1 DLCs and star$-$star DLC of an intra-night session can be written as~\citep[e.g.][]{Goyal2012A&A...544A..37G}:  

\begin{equation}
\begin{aligned}
 \label{eq.fetest}
F_{CS1}^{\eta} = \frac{\sigma^{2}_{(RQNLSy1-CS1)}} { \eta^2 \langle \sigma_{RQNLSy1-CS1}^2 \rangle}, 
\hspace{0.2cm} F_{CS2}^{\eta} = \frac{\sigma^{2}_{(RQNLSy1-CS2)}} { \eta^2 \langle \sigma_{RQNLSy1-CS2}^2 \rangle}\\ 
F_{CS1-CS2}^{\eta} = \frac{\sigma^{2}_{(CS1-CS2)}} { \eta^2 \langle \sigma_{CS1-CS2}^2 \rangle} 
\end{aligned}
\end{equation}

where $\sigma^{2}_{(RQNLSy1-CS1)}$, $\sigma^{2}_{(RQNLSy1-CS2)}$,  and $\sigma^{2}_{(CS1-CS2)}$ are the variances with $\langle \sigma_{RQNLSy1-CS1}^2 \rangle=\sum_ {i=1}^{N}\sigma^2_{i,~err}(RQNLSy1-CS1)/N$, $\langle \sigma_{RQNLSy1-CS2}^2 \rangle$, and $\langle \sigma_{CS1-CS2}^2 \rangle$ being the mean square (formal) rms errors of the i$^{th}$ data points in the DLCs of target RQNLSy1, and $N$ is the number of observations. A computed value of $\eta =1.54\pm$0.05 based upon the data of 262 intra-night monitoring sessions of AGNs by~\citet{Goyal2013JApA...34..273G} is used here for the correct use of rms errors on the photometric data points.\par
In Column 6 of online Table A2, we compare the computed $F$-values, resulting from Eq.~\ref{eq.fetest} for a session, with its estimated critical value (=$F^{(\beta)}_{cri}$) for the same session, here $\beta$ is the level of significance for the test. The $\beta$ values in the current work are set by us to be 0.05 and 0.01, corresponding to 95 percent and 99 percent confidence levels for INOV detection. The null hypothesis (i.e., non-detection of INOV) is discarded at the $\beta$ level of significance if the computed value of $F^{\eta}$ exceeds its $F^{(\beta)}_{cri}$ at the corresponding confidence level. Thus a RQNLSy1 is assigned as a variable (V) if the computed value of $F^{\eta}$ is found to be greater than its $F_{cri}(0.99)$; 
 probable variable (PV) if the same is found to be greater than $F_{cri}(0.95)$ but less or equal to $F_{cri}(0.99)$, and non-variable if $F^{\eta}$ is found to be less than or equal to $F_{cri}(0.95)$. Summary of computed values of $F^{\eta}$ and correspondingly inferred status of INOV detection for all the 28 sessions are tabulated in columns 6 and 7 of online Table A2.\par 

The second flavor of $F$-test (the $F_{p-enh}$-test) can be written following~\citet{Diego2014AJ....148...93D} as below

\begin{equation}
\label{Fenh_eq}  
\hspace{0.25cm} F_{{\rm p-enh}} = \frac{\sigma_{{\rm RQNLSy1}}^2}{\sigma_{\rm comb}^2}, \hspace{0.5cm} \sigma_{\rm comb}^2=\frac{1}{(\sum _{j=1}^q R_j) - p}\sum _{j=1}^{q}\sum _{i=1}^{R_j}D_{j,i}^2 .
\end{equation}

here $\sigma_{{\rm RQNLSy1}}^2$ is the variance of the `target RQNLSy1-reference star' DLC, while  $\sigma_{\rm comb}^2$ is  the combined variance of `comparison star-reference star' DLC having $R_{j}$ data points (number of observations) and $p$ comparison stars, computed using  scaled square deviation $D_{{\rm j, i}}^2$ as

 \begin{equation}
\hspace{2.7cm} D_{j,i}^2=\omega _j(m_{j,i}-\bar{m}_{j})^2
 \end{equation}
 
 where, $m_{j,i}$'s is the `j$^{th}$ comparison star-reference star' differential instrumental magnitudes value and $\bar{m_{j}}$  represent the corresponding average value of the DLC for its $R_{j}$ data points. The scaling factor $\omega_ {j}$ is taken here as described in~\citet{Ojha2021MNRAS.501.4110O}.\par 

 Columns 10 and 11 of online Table A2 represent the $F_{p-enh}$-test values and corresponding inferred INOV status for the entire session, following the criteria as set for $F^{\eta}$-test (see above).\par
In addition to variability resulting from two versions of the F-test, proper caution has also been taken about its PSF variation during the night before finalizing its variability status because except for one RQNLSy1 J102906.69$+$555625.2 ($z = 0.45$), all the RQNLSy1s in the present sample are at lower redshift ($\leq$ 0.4). Thus for a genuine INOV detection from the current sample, we have first carefully inspected the seeing variations of all the variable (including probable variable cases) intra-night sessions, resulting from $F_{p-enh}$-test. An RQNLSy1 is designated as V if either the FWHM of the session was fairly steady during the time of RQNLSy1’s flux variations or gradients in the FWHM of the session are anti-correlating with the systematic variations of differential magnitude of target RQNLSy1 and chosen comparison stars~\citep[see][]{Cellone2000AJ....119.1534C}.

\subsection{INOV duty cycle estimation}
\label{sec 4.1}
Adopting the definition given by ~\citet{Romero1999A&AS..135..477R}~\citep[see, also][]{Stalin2004JApA...25....1S} for the DC of intra-night variability, we have computed it for the current sample of RQNLSy1s with the following expression

\begin{equation} 
\hspace{2.5cm} DC = 100\frac{\sum_{p=1}^n C_p(1/\Delta T_p)}{\sum_{p=1}^n (1/\Delta T_p)} 
\hspace{0.1cm}{\rm per~cent} 
\label{eqno1} 
\end{equation} 

where $\Delta T_p = \Delta T_{p,~observed}(1+$z$)^{-1}$ is the observed duration of $p^{th}$ monitoring session obtained after redshift correction for the target. For $p^{th}$ session, $C_p$ is considered to be 1 in Eq.~\ref{eqno1} only when INOV is detected, otherwise $C_p$ = 0. The computed INOV duty cycles for the current sample of RQNLSy1s are listed in Table~\ref{NLSy1:DCy_result}, based on two statistical tests .\par

To compute the peak-to-peak amplitude of INOV ($\psi$) detected in a given DLC,
we followed the definition given by~\citet{Heidt1996A&A...305...42H}

\begin{equation} 
\hspace{2.5cm} \psi= \sqrt{({H_{max}}-{H_{min}})^2-2\sigma^2}
\end{equation} 

with $H_{min,~max}$ = minimum (maximum) values in the DLC of target NLSy1 relative to steady comparison stars and $\sigma^2 = \eta^2\langle\sigma^2_{NLSy1-CS}\rangle$, where, $\langle\sigma^2_{NLSy1-CS}\rangle$ is the mean square (formal) rms errors of individual data points. The mean value of ($\overline{\psi}$) for different sets (e.g., see Table~\ref{NLSy1:DCy_result}) of RQNLSy1 galaxies is computed by taking the average of the computed $\psi$ values for the DLCs belonging to the ``V'' category. In Table~\ref{NLSy1:DCy_result}, we have also summarized the DC and $\overline{\psi}$ values based on the two statistical tests for the different sets of RLNLSy1s, accessed from~\citet{Ojha2021MNRAS.501.4110O, Ojha2022MNRAS.514.5607O}.

\begin{table}
 \fontsize{7pt}{7pt}\selectfont
   \caption{Duty cycle and amplitude of INOV ($\overline{\psi}$) for the current sample of 7 RQNLSy1s based on the two versions of F-test.}
   
 \label{NLSy1:DCy_result}

 \begin{tabular}{ccccc}
    \hline
    \multicolumn{5}{ c }{Duty cycle and amplitude of INOV ($\overline{\psi}$) for the individual RQNLSy1}\\  
  \hline
    & \multicolumn{2}{ c }{$F^{\eta}$-test}                     & \multicolumn{2}{ c }{$F_{p-enh}$-test} \\
     \hline
 RQNLSy1s &       {DC}             &{$\overline{\psi}^{\dag}$}         &   {DC}             &  {$\overline{\psi}^{\dag}$} \\
          & ({\%})        &  ({\%})                           &  ({\%})            &     ({\%})          \\
  \hline
  J102906.69$+$555625.2 [3]   &       31.3 (64.3)   &   17.6 (19.5)                &   64.3 (64.3)            & 19.5 (19.5)           \\ 
  J122844.81$+$501751.3 [5]   &       22.6 (22.6)   &   32.7 (32.7)                &   38.8 (38.8)            & 26.3 (26.3)           \\ 
  J123220.11$+$495721.8 [5]   &       21.0 (21.0)   &   29.0 (29.0)                &   21.0 (42.2)            & 29.0 (21.7)           \\   
  J150916.17$+$613716.6 [3]   &       00.0 (00.0)   &   00.0 (00.0)                &   00.0 (00.0)            & 00.0 (00.0)           \\ 
  J151020.05$+$554722.0 [4]   &       00.0 (00.0)   &   00.0 (00.0)                &   25.2 (25.2)            & 10.1 (10.1)           \\ 
  J152205.41$+$393441.3 [4]   &       20.1 (20.1)   &   09.6 (09.6)                &   20.1 (40.7)            & 09.6 (07.1)           \\
  J164100.10$+$345452.7 [4]   &       00.0 (00.0)   &   00.0 (00.0)                &   29.6 (00.0)            & 11.4 (11.4)           \\ 
\hline
    \multicolumn{5}{ c }{Duty cycle and amplitude of INOV ($\overline{\psi}$) for the current and control sample of NLSy1s}\\  
  
   \hline
   &  \multicolumn{2}{ c }{$F^{\eta}$-test}                     & \multicolumn{2}{ c }{$F_{p-enh}$-test} \\
     \hline
  No. of RQNLSy1s &       {DC}             &{$\overline{\psi}^{\dag}$}         &   {DC}             &  {$\overline{\psi}^{\dag}$} \\
         &  ({\%})        &  ({\%})                           &  ({\%})            &     ({\%})          \\
  \hline
  7 RQNLSy1s [28]  &        14.3 (17.5)   &   19.2 (22.0)                          &   27.7 (35.0)            & 18.9 (17.1)           \\
   $^{\triangle}$15$\gamma$-RLNLSy1s [36]  &        30.4 (30.4)   &   14.1 (14.1)               &   40.5 (47.5)            & 13.9 (13.1)           \\
   $^{\star}$8 J-$\gamma$-RLNLSy1s [23]  &        25.9 (25.9)   &   08.6 (08.6)               &   37.5 (54.7)            & 08.1 (07.7)           \\

  \hline
  
   \multicolumn{5}{l}{$^{\dag}$The mean value for all the DLCs belonging to the type `V'. The number of sessions  }\\
   \multicolumn{5}{l}{ used is tabulated inside the bracket `[]'. $^{\bot}$Values inside parentheses have resulted }\\ 
   \multicolumn{5}{l}{when `PV' cases are considered to  `V'.}\\
    \multicolumn{5}{l}{$^{\triangle}$DC and $\overline{\psi}$ of 15 radio-loud $\gamma$-ray detected NLSy1s ($\gamma$-RLNLSy1s) are estimated }\\ 
    \multicolumn{5}{l}{using their 36 intra-night sessions from~\citet{Ojha2021MNRAS.501.4110O}.}\\
    \multicolumn{5}{l}{$^{\star}$DC and $\overline{\psi}$ of eight radio-loud jetted with $\gamma$-ray detected NLSy1s (J-$\gamma$-RLNLSy1s) }\\ 
    \multicolumn{5}{l}{are estimated using their 23 intra-night sessions from~\citet{Ojha2022MNRAS.514.5607O}.}\\

 \end{tabular}  
\end{table}

 \section{Results and discussion}
 \label{sec_4.0}
The current study presents the first attempt to systematically characterize the INOV for a sample of seven RQNLSy1s that had shown recurring flaring at 37 GHz when observed at MRO~\citep{ 2018A&A...614L...1L}. These seven NLSy1s are either radio-quiet or radio-silent (never detected in radio at any frequency). We monitored the current sample (see online Table A2) of seven RQNLSy1s in a total of 28 intra-night sessions, each $\geq$ 3 h (see online Figs. A1). It may be emphasized that an AGN may not show variability on every night it was observed, therefore to improve INOV statistics, we have devoted at least three intra-night sessions, each $\geq$ 3 h for each RQNLSy1s. 
We applied two versions of the $F-$test i.e. $F^{\eta}$-test and $F_{p-enh}$-test on the derived 28 intra-night DLCs to confirm the presence/absence of INOV in an intra-night session. Out of 28 intra-night sessions significant INOV with $\psi \gtrsim$ 10\% was detected from the DLCs of four RQNLSy1s namely J102906.69$+$555625.2, J122844.81$+$501751.3, J123220.11$+$495721.8, and J152205.41$+$393441.3 (see also first three online figures of Figs. A1) with $F^{\eta}$-test. The DLCs of another session for an RQNLSy1 J102906.69$+$555625.2 showed probable variable (PV) case with the same test even though with $\psi \gtrsim$ 21\%. The PV case would have stemmed even though $\psi \sim$ 21\%, in this case, may be due to comparatively more noise in its DLCs (see column 12 of online Table A2). Furthermore, eight variable cases with $\psi \gtrsim$ 10\% from the DLCs of six RQNLSy1s resulted with $F_{p-enh}$-test. Another two sessions with $\psi > $ 4\% were placed under PV category using the same test. Thus, using statistical tests (see online table Table A2) we find that all RQNLSy1s exhibit INOV except one RQNLSy1, J150916.17$+$613716.6 that did not show INOV with any of the statistical tests for all intra-night sessions. It may be recalled here that out of seven RQNLSy1s observed with JVLA, RQNLSy1s J150916.17$+$613716.6 is the only one with absolutely no detection in the JVLA, while the others are typically showing at least one data point at some frequencies~\citep[see][]{Berton2020A&A...636A..64B, 2023arXiv:2312.02326J}. Therefore, the non-detection of INOV in this source may be due n to its quiescent phase in the optical band too. \par
From Table~\ref{NLSy1:DCy_result}, a DC of $\sim$ 28\% ($\sim$ 35\% when two `PV' cases are considered to be `V') with $\overline{\psi}$ (the mean value of $\psi$ for all the DLCs belonging to the type `V') of $\sim$ 19\% resulted from $F_{p-enh}$-test for the current sample. However, DC of $\sim$ 14\% ($\sim$ 18\% when a PV case is also considered to be `V') with $\overline{\psi}$ of 19\% are estimated based on the more conservative $F^{\eta}$-test. Considering the average DC ($\overline{DC}$) of conservative $F^{\eta}$-test and $F_{p-enh}$-test which is $\sim$ 21\% ($\sim$ 26\% when `PV' cases are considered to be `V), found to be comparable to DC of 25\% $-$ 30\%, exhibited by $\gamma$-RLNLSy1s, that display blazar-like INOV~\citep{Ojha2021MNRAS.501.4110O}.
The strong INOV level of $\psi \gtrsim$ 10\% in any variable cases resulting from the current sample appear striking because even powerful radio-quiet quasars empowered by SMBH never displayed $\psi >$ 4\%~\citep[e.g., see][and references therein]{Gopal-Krishna2018BSRSL..87..281G}. From Table~\ref{NLSy1:DCy_result}, it also appears that duty cycles of individual variable sources are always  $\gtrsim$ 20\%  with $\psi \gtrsim$ 10\%, consistent with the DC estimated for the whole sample. Here, it may be recalled that in the extensive INOV study of six prominent classes of luminous AGNs covered in 262 monitoring sessions by~\citet{Goyal2013MNRAS.435.1300G} where a very similar telescopes and analysis procedure were used, only blazars class of AGN displayed a DC above $\sim$ 10\% for  $\psi >$ 3\%. Thus, we conclude that the DC resulting from the present sample appears to be blazar-like.\par
Considering the median black hole mass log($M_{BH}/M_{\sun}$) = 6.84 that implied to harbor inevitably less powerful jets for the current sample~\citep{Heinz2003MNRAS.343L..59H, Foschini2014IJMPS..2860188F}. The resulting $\overline{DC}$ of 21\% from the current sample appears striking and it suggests the origin of their INOV from the relativistic plasma jets. However, it may be recalled here that the median black hole mass of the current sample is an order of less massive than the median black hole mass of log(M$_{BH}/M_{\sun}$) = 7.72 of jetted-RLNLSy1s ~\citep[see last column of Table 5 of][]{Ojha2022MNRAS.514.5607O}. Thus, despite having low SMBH our sample of NLSy1s is capable of launching relativistic jets. This is in agreement with the recent results of blazar-like INOV displayed by a sample of 12 low-mass (median M$_{BH} = 10^{6} M_{\sun}$) radio-quiet AGNs~\citep{Gopal-Krishna2023MNRAS.518L..13G}, which strengthened the case for the ability to launch relativistic jets from apparently radio-quiet AGNs.\par
The detection of recurring flaring at 37 GHz at MRO~\citep{2018A&A...614L...1L} and strong INOV levels found here from the current sample of seven RQNLSy1s hints at the presence of relativistic jets in the current sample of RQNLSy1s, however, non-detection of jet activity in JVLA observations at 1.6, 5.2, and 9.0 GHz frequencies appears contrasting. Here, it may be recalled that powerful relativistic jets that are capable of propagating outside the host galaxy are found in approximately 10\% of AGNs~\citep{Padovani2017NatAs...1E.194P}. Therefore, non-detection of jet activity in JVLA observations from the current sample may be due to their low integrated luminosity which is indeed low $\leq$ 1.0$\times 10^{39}$ erg s$^{-1}$~\citep[see Table 2 of][]{Berton2020A&A...636A..64B}. Such scenario straightens with integrated luminosity $\geq$ 1.5$\times 10^{39}$ erg s$^{-1}$ found in jetted sources~\citep[see][]{Berton2018A&A...614A..87B}. In addition to the low power of jets, non-detection of jet activity in JVLA observations might be due to absorbed jets because of a more tied connection of optical emission to nuclear jet emission at millimeter wavelengths as compared to its emission at lower radio frequencies~\citep[see][]{Gopal-Krishna2023MNRAS.518L..13G}, which is largely attenuated due to a high opacity around the nuclear jet, as interpreted from Very Long Baseline Interferometry studies~\citep{Gopal-Krishna1991vagn.conf..194G, Boccardi2017A&ARv..25....4B}.\par 
The detection of multiple flaring at 37 GHz and INOV in the current work while non-detection of jet activity in JVLA observations at 1.6, 5.2, and 9.0 GHz frequencies may be due to the flaring and quiescent states of RQNLSy1s when 37 GHZ, our observations, and JVLA observations were taken place, respectively.\par
Another possibility of non-detection of jet activity in JVLA observations as also emphasised in~\citet{Berton2020A&A...636A..64B} from the current sample could be due to their high-frequency peakers nature which usually happens with extremely young objects like NLSy1s which are expected to be in an early phase of their evolution~\citep[see][]{Komossa2018rnls.confE..15K, Paliya2019JApA...40...39P}. It may be emphasised here that NLSy1s are young and typically characterized by high Eddington ratios~\citep{Boroson1992ApJS...80..109B, Ojha2020ApJ...896...95O}, therefore expected to be associated with a dense circumnuclear environment around non-flattened broad-line region~\citep{Heckman2014ARA&A..52..589H, Vietri2018rnls.confE..47V, Berton2020CoSka..50..270B} that might also hinder relativistic jets propagation through its interaction with the clouds~\citep{vanBreugel1984AJ.....89....5V}.\par
In addition to non-detection of jet activity in the JVLA observations at 1.6, 5.2, and 9.0 GHz, recent observations with JVLA at higher frequencies 10, 15, 22, 33, and, 45 GHz showed no signs of jets from the current sample rather than resulting in either steep spectrum or no detection at all from most of the sources except for a source  J122844.81$+$501751.2 that showed the flat spectrum~\citep{2023arXiv:2312.02326J}. However, blazar-like INOV found here from the current sample of RQNLSy1s could be due to the peculiar geometry of jet in these sources that causes changes in viewing angle toward the observer's line of sight thus changes in Doppler factor~\citep{Raiteri2017Natur.552..374R}.\par 
Finally, one last possibility of non-detection of jet activity in JVLA observations might be magnetic reconnection in the magnetosphere of black hole~\citep{Lazarian1999ApJ...517..700L, 2010A&A...518A...5D, Kadowaki2015ApJ...802..113K, Ripperda2022ApJ...924L..32R, Kimura2022ApJ...937L..34K} which does not require the presence of a permanent relativistic jet and can account timescales of variability ranging from minutes to days. In brief, this model invokes the interactions of magnetic field lines emerging from the accretion disk with the magnetosphere anchored into the central black hole horizon~\citep{Blandford1977MNRAS.179..433B}. With the enhancement of the accretion rate, magnetic fluxes from the accretion disc and those anchored into the black hole horizon are pushed together in the inner disk region and reconnected under finite magnetic resistivity~\citep[see Figure 1 of][]{Kadowaki2015ApJ...802..113K}. This reconnection becomes very efficient and fast under turbulence instability and releases a huge amount of magnetic power. A part of released magnetic power accelerates particles to relativistic velocities and thus is attributed to radio emissions and flares. Therefore, recurring flaring at 37 GHz observed at MRO might be due to fast magnetic reconnection driven by turbulence. Thus, it implies that jets may not be present in our sample as evidenced from their recent radio studies by JVLA and VLBA~\citep[see][]{Berton2020CoSka..50..270B, 2023arXiv:2312.02326J}, but the INOV seen in the current study may be due to magnetic reconnection in the black hole magnetosphere. Furthermore, for the low luminosity AGNs, ~\citet{2010A&A...518A...5D} and ~\citet{Kadowaki2015ApJ...802..113K} showed that the turbulence-induced fast magnetic reconnection events in an efficiently accreting black hole with a mass of M$_{BH} \sim 10^{6} M_{\sun}$ can release $\sim10^{39}$ to $10^{43}$ erg s$^{-1}$ power which is sufficient to explain the detection of INOV and flares. The jetted AGN such as gamma-ray NLSy1s and blazars can also possess magnetic reconnection and instabilities in their accretion disks, however, jet-dominated non-thermal emission completely overwhelms the thermal emission from the accretion disc in them~\citep[see][]{Mangalam1993ApJ...406..420M, Ulrich1997ARA&A..35..445U, Blandford2019ARA&A..57..467B}.

Additionally, magnetic reconnection events causing acceleration of plasma to relativistic velocities can also potentially heat the corona of the accretion disk resulting in the enhancement of thermal X-ray emission. Furthermore, emissions in high energy gamma-rays might also be possible through interactions of these accelerated relativistic electrons with photon density in the surrounds of the black hole via SSC and/or IC interactions~\citep{Kadowaki2015ApJ...802..113K}.  Therefore, coordinated radio, X-ray, and gamma-ray observations can be useful in confirming or ruling out the possibility of a magnetic reconnection mechanism~\citep{2010A&A...518A...5D, Kimura2022ApJ...937L..34K}. \par

\section{Conclusions}
In the current study, we present the first attempt to systematically characterize the INOV for a sample of seven RQNLSy1s that had shown recurring flaring at 37 GHz when observed with MRO~\citep{2017A&A...603A.100L, 2018A&A...614L...1L}, however, no signs of jet were detected from them when these RQNLSy1s were observed in radio bands with JVLA at low frequencies.  Recurring flaring at 37 GHz from them strongly suggests the presence of jets in them. However, the non-detection of jet activity in JVLA and VLBA observations appears puzzling. We have addressed this issue by taking advantage of INOV which is being used to infer the presence of relativistic jets in AGNs based on their blazar-like duty cycle and the amplitude of INOV. Therefore, we monitored the current sample in a total of 28 intra-night sessions, each $\geq$ 3 h. The resulting level of INOV from this sample found to be similar to the INOV level of $\gamma$-RLNLSy1s, which displays blazar-like INOV. Thus detection of recurring flaring at 37 GHz with MRO and strong INOV level found here from the current sample of seven RQNLSy1s hints at the presence of relativistic jets in the present set of RQNLSy1s. Furthermore, the resulting strong level of INOV from the current sample of low-mass RQNLSy1s and almost a similar INOV level observed recently by~\citet{Gopal-Krishna2023MNRAS.518L..13G} from 12 low-mass active galactic nuclei, suggest that even low-mass radio-quiet and/or radio-silent AGNs can launch relativistic jets. Furthermore, inferred jet activity from the current sample along with the presence of jet activity in 12 low-mass AGNs~\citep{Gopal-Krishna2023MNRAS.518L..13G} would be useful in understanding relativistic jet mechanism in lower black hole mass (M$_{BH} \sim 10^{6}- 10^{7} M_{\sun}$) AGNs.\par
The resulting variability nature of these sources in radio and optical wavebands is difficult to explain with the usual variability mechanisms of AGNs. Therefore, studying such types of sources with bigger sample sizes is very important to unveil the nature of these sources as they might represent a new type of AGN variability. 

\section*{Acknowledgements}
We acknowledge the Director of Physical Research Laboratory (PRL) and the Department of Space (DOS), Government of India, for supporting this research work and VO's postdoctoral fellowship. The PRL runs and supports the Observational activities with the 1.2m and 2.5m telescopes at Mt. Abu. We acknowledge the PRL Mt. Abu Observatory staff for our observations with the PRL's 1.2m and 2.5m telescopes. This work has used observations from the Mt. Abu Faint Object Spectrograph and Camera-Pathfinder (MFOSC-P) and we acknowledge the help from Dr. Vipin Kumar and the entire MFOSC-P team led by Dr. Mudit K. Srivastava. This work has also used observations with the Faint Object Camera (FOC) on the 2.5m telescope and we acknowledge the efforts of the FOC development team led by Prof. Abhijit Chakraborty and Dr. Vishal Joshi. We also thank the Aryabhatta Research Institute of Observational Sciences (ARIES) authorities and staff for their assistance during the observations taken from the 1.04m Sampurnanand Telescope (ST) at Nainital, 1.30m Devasthal Fast Optical Telescope (DFOT) at Devasthal, both ST and DFOT are run by ARIES. We thank the anonymous referee for providing useful comments and suggestions, which helped us to improve the manuscript.

 \section*{Data availability}
The data from the \href{https://www.aries.res.in/facilities/astronomical-telescopes}{ARIES telescopes} and the Mount Abu observatory \href{https://www.prl.res.in/~miro/telescopes.html}{Mt. Abu telescopes of PRL} used in this paper will be shared on reasonable request to the corresponding author.


\label{lastpage}


\appendix

\section{Appendix}

\begin{table*}
  \caption{Basic observational parameters and log of the target RQNLSy1s and comparison stars used in the current study. Columns are listed as follows: (1) RQNLSy1s and the comparison stars; (2) date(s) of monitoring; (3) right ascension (RA.); (4) declination (DEC.); (5) SDSS $g$-band magnitude; (6) SDSS $r$-band magnitude; (7) SDSS `$g-r$' colours. The positions and apparent magnitudes of the sources and their comparison stars were taken from the SDSS DR14~\citep{Abolfathi2018ApJS..235...42A}. `B-R' colours have been used from USNO-A2.0 catalog~\citep{Monet1998AAS...19312003M} for the target RQNLSy1s J152205.41$+$393441.3 and its comparison stars (marked by `$\star$') due to non-availability of its SDSS `$g-r$' colours.}

  \label{tab_gray_comp_star}
\begin{tabular}{ccc ccc c}\\
\hline
{RQNLSy1s and} &   Date(s) of monitoring       &   {RA.(J2000)} & {DEC.(J2000)}              & {\it g} & {\it r} & \hspace{0.1 cm} {\it g-r} \\
the comparison stars           &      &   (hh mm ss)       &(dd mm ss)   & (mag)   & (mag)   & (mag)     \\
{(1)}      & {(2)}        & {(3)}           & {(4)}                              & {(5)}   & {(6)}   & \hspace{0.1 cm} {(7)}     \\
\hline
\multicolumn{7}{l}{}\\
J102906.69$+$555625.2 & 2020 Dec. 19; 2021 Apr. 12; 2022 Dec. 27                 & 10:29:06.69  &$+$55:56:25.22 & 19.08 & 19.07& 0.01  \\
S1                    & 2020 Dec. 19; 2022 Dec. 27                               & 10:28:41.50  &$+$55:57:21.89 & 19.69 & 18.34& 1.35  \\
S2                    & 2020 Dec. 19; 2022 Dec. 27                               & 10:28:21.48  &$+$55:58:34.46 & 19.38 & 18.18& 1.20  \\
S3                    & 2021 Apr. 12                                             & 10:29:57.00  &$+$55:54:27.70 & 18.59 & 17.67& 0.92  \\
S4                    & 2021 Apr. 12                                             & 10:29:54.51  &$+$55:55:04.26 & 19.25 & 17.82& 1.43  \\
J122844.81$+$501751.3 & 2020 Dec. 23; 2021 Apr. 10; 2022 Mar. 27, Apr. 09, May 08 & 12:28:44.81  &$+$50:17:51.27 & 18.44 & 17.88& 0.56 \\
S1                    & 2020 Dec. 23                                             & 12:28:20.67  &$+$50:22:08.95 & 18.04 & 17.28& 0.76  \\
S2                    & 2020 Dec. 23                                             & 12:28:58.53  &$+$50:20:11.89 & 16.55 & 15.75& 0.80  \\
S3                    & 2021 Apr. 10                                             & 12:28:34.32  &$+$50:15:40.12 & 17.10 & 16.30& 0.80  \\
S4                    & 2021 Apr. 10, 2022 Apr. 09                               & 12:28:48.49  &$+$50:15:26.72 & 17.46 & 16.60& 0.86  \\
S5                    & 2022 Mar. 27, May 08                                     & 12:28:54.32  &$+$50:16:22.36 & 18.64 & 17.39& 1.25  \\
S6                    & 2022 Mar. 27, May 08                                     & 12:28:45.45  &$+$50:14:48.60 & 17.58 & 17.21& 0.37  \\
S7                    & 2022 Apr. 09                                             & 12:28:45.32  &$+$50:18:40.78 & 18.11 & 17.49& 0.73  \\
J123220.11$+$495721.8 & 2021 Jan. 24, Apr. 11; 2022 Jan 14, 23, 24               & 12:32:20.11  &$+$49:57:21.79 & 17.69 & 17.54& 0.62  \\
S1                    & 2021 Jan. 24, Apr. 11; 2022 Jan 14                       & 12:31:59.11  &$+$49:58:50.29 & 17.65 & 16.24& 1.41  \\
S2                    & 2021 Jan. 24                                             & 12:32:55.82  &$+$49:52:58.96 & 18.08 & 17.23& 0.85  \\
S3                    & 2021 Apr. 11; 2022 Jan. 14, 23, 24                       & 12:32:15.53  &$+$49:56:07.70 & 17.15 & 15.93& 1.22  \\
S4                    & 2022 Jan 23, 24                                          & 12:32:09.86  &$+$49:56:08.93 & 17.35 & 17.03& 0.32  \\
J150916.17$+$613716.6 & 2021 Apr. 08, 10; 2022 Mar. 27                           & 15:09:16.17  &$+$61:37:16.75 & 18.88 & 18.47& 0.41  \\
S1                    & 2021 Apr. 08, 10                                         & 15:09:07.14  &$+$61:37:45.68 & 17.41 & 16.96& 0.45  \\
S2                    & 2021 Apr. 08, 10                                         & 15:09:32.71  &$+$61:37:57.88 & 17.71 & 16.66& 1.05  \\
S3                    & 2022 Mar. 27                                             & 15:09:38.99  &$+$61:34:36.71 & 18.56 & 17.42& 1.14  \\
S4                    & 2022 Mar. 27                                             & 15:09:19.99  &$+$61:32:17.34 & 16.80 & 16.43& 0.37  \\
J151020.05$+$554722.0 & 2021 Apr. 07, 11; 2022 Apr. 10, 23                       & 15:10:20.05  &$+$55:47:22.05 & 18.46 & 17.77& 0.69  \\
S1                    & 2021 Apr. 07                                             & 15:09:44.54  &$+$55:44:07.17 & 18.27 & 17.27& 1.00  \\
S2                    & 2021 Apr. 07, 11                                         & 15:09:59.61  &$+$55:51:28.95 & 17.38 & 16.98& 0.40  \\
S3                    & 2021 Apr. 11; 2022 Apr. 10                               & 15:10:02.94  &$+$55:49:18.50 & 16.95 & 15.91& 1.04  \\
S4                    & 2021 Apr. 11; 2022 Apr. 10, 23                           & 15:10:17.66  &$+$55:47:41.71 & 17.65 & 16.72& 0.93  \\
S5                    & 2022 Apr. 23                                             & 15:10:17.29  &$+$55:50:47.03 & 15.90 & 15.49& 0.41  \\
J152205.41$+$393441.3 & 2021 Jun. 01, 02, 03; 2022 Feb. 21                       & 15 22 05.41  &$+$39 34 41.30 & 14.90${^\star}$ & 13.10$^{\star}$& 1.80${^\star}$\\
S1                    & 2021 Jun. 01, 02                                         & 15:22:15.47  &$+$39:30:41.30 & 15.50${^\star}$ & 15.49${^\star}$& 0.01${^\star}$ \\
S2                    & 2021 Jun. 01, 02                                         & 15:22:15.07  &$+$39:36:13.70 & 16.90${^\star}$ & 15.50${^\star}$& 1.40${^\star}$ \\
S3                    & 2021 Jun. 03                                             & 15:21:22.87  &$+$39:35:07.20 & 15.80${^\star}$ & 15.20${^\star}$& 0.60${^\star}$ \\
S4                    & 2021 Jun. 03                                             & 15:21:18.16  &$+$39:35:49.60 & 15.50${^\star}$ & 15.20${^\star}$& 0.30${^\star}$ \\
S5                    & 2022 Feb. 21                                             & 15:22:02.09  &$+$39:31:31.10 & 15.70${^\star}$ & 15.69${^\star}$& 0.01${^\star}$ \\
S6                    & 2022 Feb. 21                                             & 15:21:53.28  &$+$39:32:35.40 & 14.30${^\star}$ & 14.20${^\star}$& 0.10${^\star}$ \\
J164100.10$+$345452.7 & 2021 June, 01, 06; 2022 Apr. 23, May 07                  & 16:41:00.10  &$+$34:54:52.68 & 17.94 & 16.96& 0.98 \\
S1                    & 2021 June, 01                                            & 16:41:07.53  &$+$34:49:34.97 & 17.26 & 16.09& 1.17  \\
S2                    & 2021 June, 01, 06                                        & 16:40:39.77  &$+$34:57:31.21 & 16.96 & 15.98& 0.98  \\
S3                    & 2021 June, 06                                            & 16:40:37.99  &$+$35:00:33.55 & 16.21 & 15.64& 0.57  \\
S4                    & 2022 Apr. 23, May 07                                     & 16:41:10.04  &$+$34:51:09.22 & 16.47 & 16.10& 0.37  \\
S5                    & 2022 Apr. 23, May 07                                     & 16:41:00.67  &$+$34:53:31.39 & 17.37 & 15.98& 1.39  \\

\hline

\end{tabular}
\end{table*}

\begin{table*}
  \begin{minipage}{195mm}
 \begin{center}   
 {\small
  \caption[caption]{Observation dates, duration of monitoring along with DLCs details, and the status of the statistical tests for the sample of 7 RQNLSy1 galaxies studied in the present work.}
  \label{NLSy1:tab_result} \begin{tabular}{cccc ccccc ccccl}
  \hline
  {RQNLSy1s} & Date(s)$^{a}$ &  T$^{b}$  & N$^{c}$  & Median$^{d}$ & {$F^{\eta}$-test} & {INOV}  & { {$F^{\eta}$-test}}& { Variability} & {$F_{p-enh}$-test} & {INOV} &{$\sqrt { \langle \sigma^2_{i,err} \rangle}$} & $\overline\psi^{g}_{s1, s2}$& \\
  (SDSS name) & yyyy.mm.dd & (hrs) & & FWHM  & {$F_{s1}^{\eta}$},{$F_{s2}^{\eta}$} & status$^{e}$ & { {$F_{s1-s2}^{\eta}$}}& { status of}  & $F_{p-enh}$ & status$^{f}$  & (AGN-s)$^{g}$ & (\%) &\\
  &&&& (arcsec) &           &{99\%}& { 99\%}& { (s1$-$s2)$^{e}$} &  &{99\%} & &\\
  {(1)}&{(2)} & {(3)} & {(4)} & {(5)} & {(6)} & {(7)} & {(8)} & {(9)} & {(10)} & {(11)} &  {(12)}&  {(13)} &\\
\hline
                                             
 J102906.69$+$555625.2 & 2020.12.19 & 4.80 & 17& 2.56& 03.40, 03.41 &  V,  V & 00.86 & NV & 03.64 &  V & 0.03 & 17.62&\\  
                       & 2021.04.12 & 4.55 & 14& 2.22& 03.06, 02.62 & PV, PV & 00.77 & NV & 03.96 &  V & 0.04 & 21.33&\\
                       & 2022.12.27 & 4.21 & 22& 2.27& 00.16, 00.14 & NV, NV & 00.29 & NV & 00.53 & NV & 0.05 &    --&\\
 J122844.81$+$501751.3 & 2020.12.23 & 3.49 & 22& 3.53& 00.71, 01.09 & NV, NV & 00.81 & NV & 00.88 & NV & 0.02 &    --&\\
                       & 2021.04.10 & 3.10 & 36& 2.30& 00.61, 00.53 & NV, NV & 00.64 & NV & 00.96 & NV & 0.04 &    --&\\ 
                       & 2022.03.27 & 4.49 & 59& 1.79& 00.44, 00.34 & NV, NV & 00.35 & NV & 01.25 & NV & 0.04 &    --&\\
                       & 2022.04.09 & 3.26 & 32& 1.87& 02.38, 02.40 &  V,  V & 00.67 & NV & 03.21 &  V & 0.05 & 32.67&\\
                       & 2022.05.08 & 4.53 & 17& 1.90& 00.59, 00.55 & NV, NV & 00.14 & NV & 04.37 &  V & 0.07 & 19.99&\\
 J123220.11$+$495721.8 & 2021.01.24 & 3.25 & 19& 1.86& 00.71, 00.50 & NV, NV & 00.41 & NV & 01.73 & NV & 0.01 &    --&\\
                       & 2021.04.11 & 3.69 & 43& 2.33& 00.35, 00.34 & NV, NV & 00.57 & NV & 00.61 & NV & 0.02 &    --&\\
                       & 2022.01.14 & 3.04 & 15& 1.92& 02.23, 01.90 & NV, NV & 02.10 & NV & 01.06 & NV & 0.02 &   - -&\\
                       & 2022.01.23 & 3.03 & 17& 2.44& 05.44, 05.54 &  V, V  & 00.87 & NV & 06.27 & V  & 0.03 & 28.96&\\
                       & 2022.01.24 & 3.00 & 30& 2.18& 01.50, 01.62 & NV, NV & 00.68 & NV & 02.20 & PV & 0.03 & (14.36)&\\
 J150916.17$+$613716.6 & 2021.04.08 & 3.20 & 13& 2.71& 01.83, 01.92 & NV, NV & 01.84 & NV & 00.99 & NV & 0.03 &    --&\\
                       & 2021.04.10 & 4.02 & 16& 2.18& 00.93, 00.95 & NV, NV & 00.52 & NV & 01.79 & NV & 0.03 &    --&\\
                       & 2022.03.27 & 3.11 & 12& 2.10& 01.12, 01.50 & NV, NV & 01.11 & NV & 01.01 & NV & 0.04 &    --&\\
 J151020.05$+$554722.0 & 2021.04.07 & 3.14 & 18& 1.86& 00.71, 00.84 & NV, NV & 00.56 & NV & 01.28 & NV & 0.02 &    --&\\
                       & 2021.04.11 & 3.45 & 14& 2.44& 02.25, 02.35 & NV, NV & 00.46 & NV & 04.85 & V  & 0.02 & 10.13&\\
                       & 2022.04.10 & 4.16 & 11& 2.40& 00.44, 00.61 & NV, NV & 00.99 & NV & 00.44 & NV & 0.03 &    --&\\
                       & 2022.04.23 & 3.30 & 10& 2.41& 00.95, 00.33 & NV, NV & 01.22 & NV & 00.77 & NV & 0.04 &    --&\\
 J152205.41$+$393441.3 & 2021.06.01 & 3.10 & 36& 2.68& 00.93, 01.10 & NV, NV & 00.68 & NV & 01.38 & NV & 0.01 &    --&\\
      		     & 2021.06.02 & 3.81 & 35& 3.06& 03.65, 04.78 &  V,  V & 01.14 & NV & 03.21 & V  & 0.01 & 09.60&\\
          	     & 2021.06.03 & 3.70 & 41& 2.70& 01.52, 01.49 & NV, NV & 00.81 & NV & 01.89 & PV & 0.01 & (04.57)&\\
 		         & 2022.02.21 & 2.21 & 23& 2.48& 01.16, 02.34 & NV, NV & 02.02 & NV & 00.51 & NV & 0.01 &    --&\\
 J164100.10$+$345452.7 & 2021.06.01 & 3.00 & 29& 3.16& 00.87, 00.74 & NV, NV & 00.49 & NV & 01.78 & NV & 0.02 &    --&\\
     		       & 2021.06.06 & 6.95 & 67& 2.82& 00.77, 00.66 & NV, NV & 00.63 & NV & 01.22 & NV & 0.02 &    --&\\
 		         & 2022.04.23 & 3.23 & 13& 2.30& 01.64, 02.09 & NV, NV & 01.56 & NV & 01.05 & NV & 0.01 &    --&\\
 		           & 2022.05.07 & 3.02 & 10& 2.54& 03.14, 02.92 & NV, NV & 00.35 & NV & 08.94 &  V & 0.02 & 11.45 &\\							
 \hline
  \multicolumn{13}{l}{$^{a}$Date(s) of the monitoring session(s). $^{b}$Duration of the monitoring session in the observed frame. $^{c}$Number of data points in the DLCs of the monitoring session.}\\ 
  \multicolumn{13}{l}{$^{d}$Median seeing (FWHM in arcsec) for the session. $^{e,~f}$Variability status inferred from F$^{\eta}$ and F$_{p-enh}$ tests, with V = variable, i.e. confidence level $\geq$ 99\%;}\\
  \multicolumn{13}{l}{PV = probable variable, i.e. $95-99$\% confidence level; NV = non-variable, i.e. confidence level $<$ 95\%.}\\
\multicolumn{13}{l}{$^{g}$Mean amplitude of variability in the two DLCs of the target RQNLSy1 (i.e., relative to the two chosen comparison stars).}\\
    \end{tabular}  
 }              
\end{center}
 \end{minipage} 
    \end{table*}

\begin{figure*}
\centering
\includegraphics[height=21cm, width=19cm]{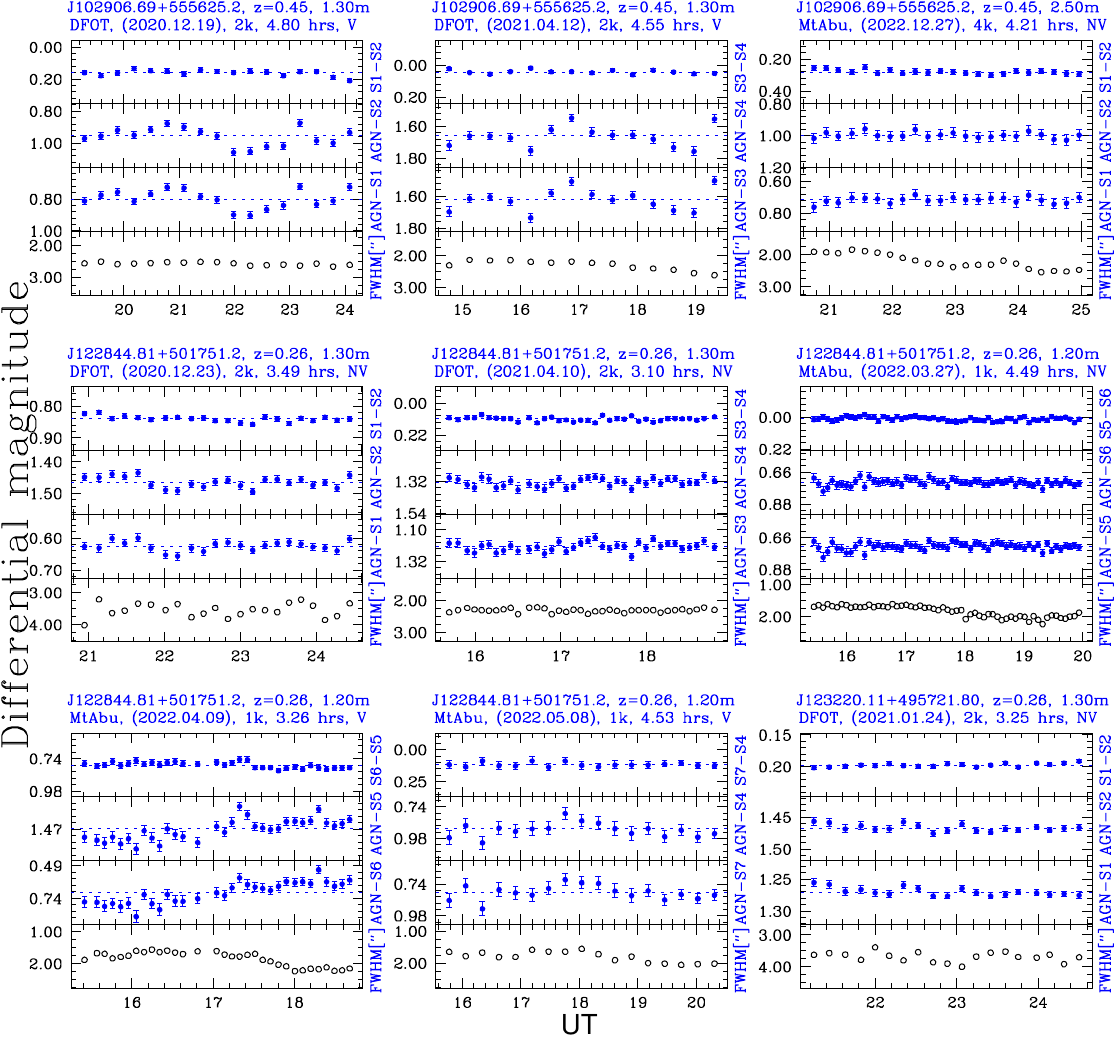}
\caption[]{Differential light curves (DLCs) of three RQNLSy1s from the sample of seven RQNLSy1s. Top of each panel shows target name, its redshift, and a few observational details along with its variability status. In each panel, DLCs from top to bottom are for two chosen non-variable comparison stars ($S_{i}-S_{j}$), target RQNLSy1, comparison star ($RQNLSy1-S_{j}$), and target RQNLSy1, reference star ($RQNLSy1-S_{i}$), respectively, while seeing (FWHM in arcseconds) variation of the session is shown in the bottom panel.}
\addtocounter{figure}{-1}
\label{fig:lurve 1}
\end{figure*}

\begin{figure*}
\centering
\includegraphics[height=21cm, width=19cm]{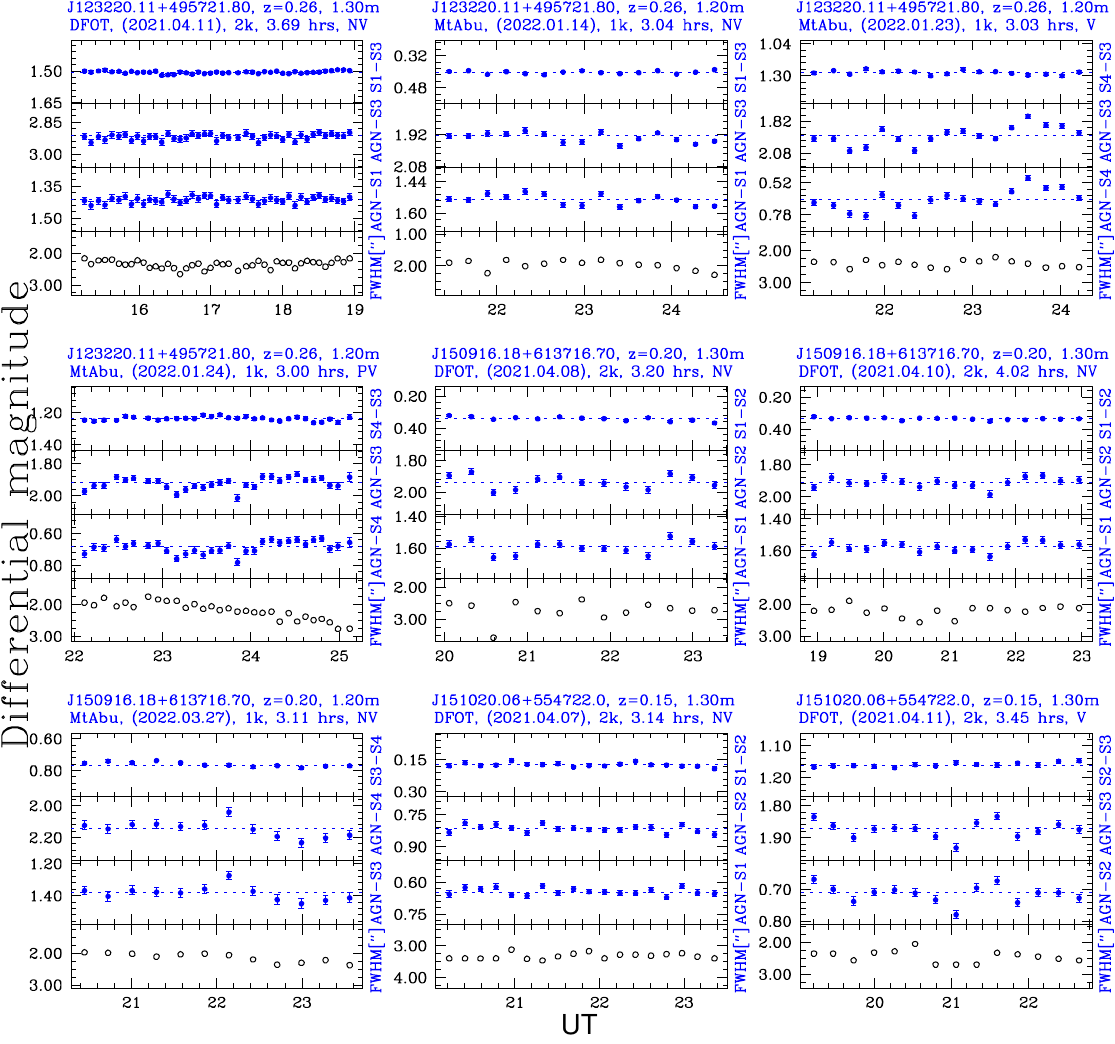}
\caption[]{(continued) DLCs for the subsequent two RQNLSy1s from the present sample of seven RQNLSy1s.}
\addtocounter{figure}{-1}
\label{fig:lurve 2}
\end{figure*}

\begin{figure*}
\centering
\includegraphics[height=21cm, width=19cm]{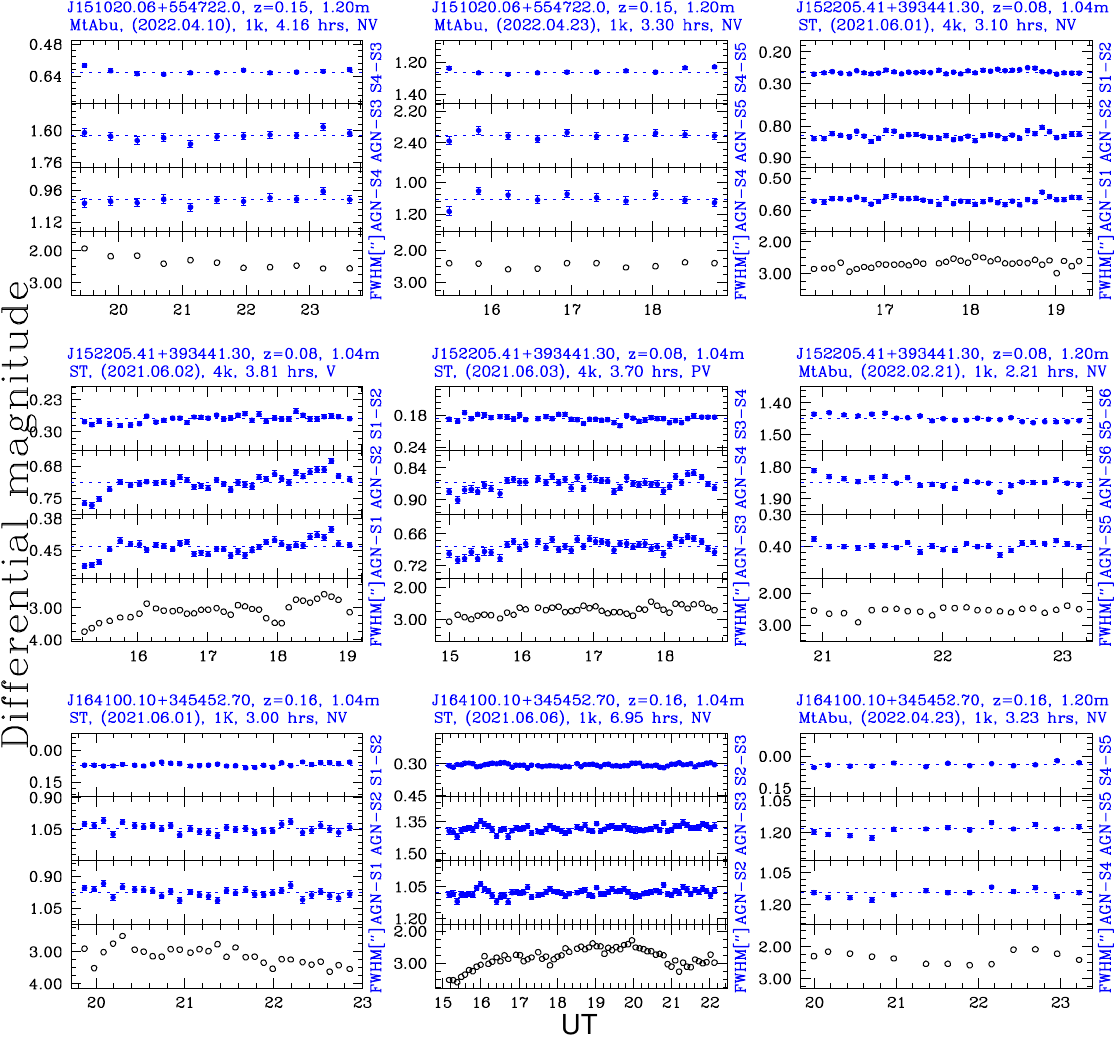}
\caption[]{(continued) DLCs for last two RQNLSy1s from the present sample of seven RQNLSy1s.}
\addtocounter{figure}{-1}
\label{fig:lurve 3}
\end{figure*}

\begin{figure}
	\begin{center}
	\includegraphics[width=8.0cm, height=7.4cm]{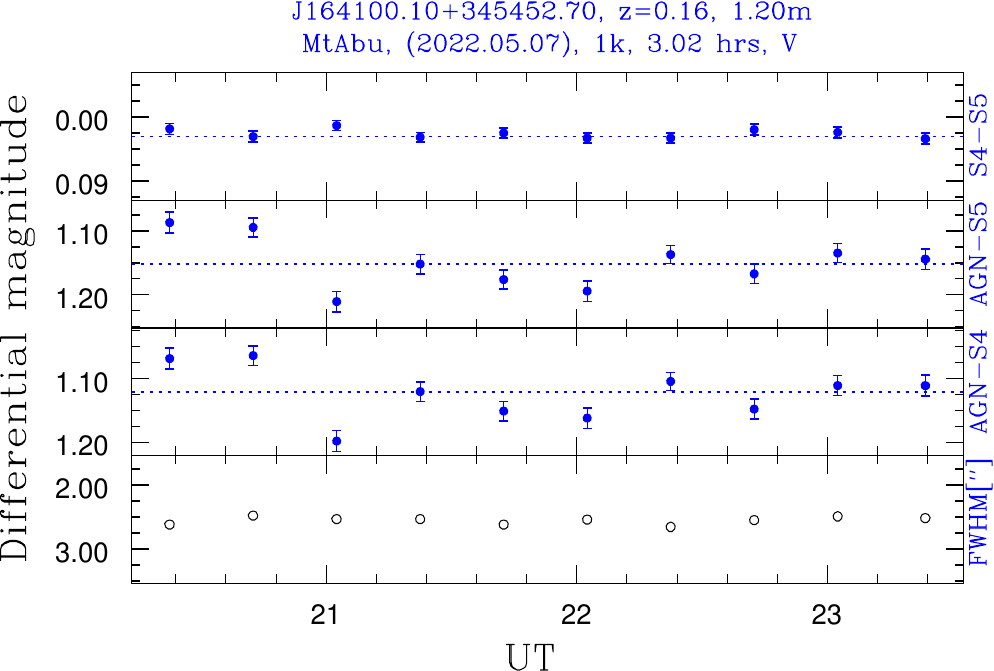}
	\end{center}
 	\caption[]{(continued) last DLC of RQNLSy1 J164100.10$+$345452.7 from the present sample of seven RQNLSy1s.}
	\label{fig:lurve 4}
 \addtocounter{figure}{-1}
\end{figure}

 \end{document}